# STABLE EQUILIBRIA OF ANISOTROPIC PARTICLES ON SUBSTRATES: A GENERALIZED WINTERBOTTOM CONSTRUCTION

WEIZHU BAO[*], WEI JIANG[†], DAVID J. SROLOVITZ[‡], AND YAN WANG[§]

**Abstract.** We present a new approach for predicting stable equilibrium shapes of two-dimensional crystalline islands on flat substrates, as commonly occur through solid-state dewetting of thin films. The new theory is a generalization of the widely used Winterbottom construction (*i.e.*, an extension of the Wulff construction for particles on substrates). This approach is equally applicable to cases where the crystal surface energy is isotropic, weakly anisotropic, strongly anisotropic and "cusped". We demonstrate that, unlike in the classical Winterbottom approach, multiple equilibrium island shapes may be possible when the surface energy is strongly anisotropic. We analyze these shapes through perturbation analysis, by calculating the first and second variations of the total free energy functional with respect to contact locations and island shape. Based on this analysis, we find the necessary conditions for the equilibria to be stable to two-dimensional perturbations and exploit this through a generalization of the Winterbottom construction to identify all possible stable equilibrium shapes. Finally, we propose a dynamical evolution method based on surface diffusion mass transport to determine whether all of the stable equilibrium shapes are dynamically accessible. Applying this approach, we demonstrate that islands with different initial shapes may evolve into different stationary shapes and show that these dynamically-determined stationary states correspond to the predicted stable equilibrium shapes, as obtained from the generalized Winterbottom construction.

**1. Introduction.** Many micro- and nano-scale devices use solid thin films as basic building components. Compared to traditional bulk materials, thin films have a very large surface-area-to-volume ratios. Such films can be unstable to particle formation (dewetting or agglomeration) due to surface tension/capillarity effects, especially at high temperatures and/or on long time scales. Solid-state dewetting has been observed and studied in a large number of experimental systems (e.g., see the recent review papers [26, 39]), such as Ni films on MgO substrates and Si films on amorphous $SiO_2$ substrates (*i.e.*, SOI).

The dewetting of solid thin films is similar to the dewetting of liquid films [3, 35, 46]. The main difference between solid and liquid dewetting is associated with anisotropy of physical properties (e.g., surface energy, diffusivity) and the mode of mass transport; solid-state dewetting is usually dominated by surface diffusion rather than fluid dynamics (as in liquid-state dewetting). Solid-state dewetting is increasingly important in many modern technologies where it can be either detrimental or advantageous. For example, dewetting can destroy micro-/nano-device perfor-mance through instabilities in carefully patterned structures (this is a major reliability issue). On the other hand, dewetting of a continuous film can be used to create patterns of nanoscale particles which can be used for sensors [30], optical and magnetic devices [1], and as catalysts for the growth of carbon or semiconductor nanowires [36]. Therefore, solid-state dewetting is attracting ever more attention (e.g., see [5, 15, 17, 33, 34, 37, 44, 50]).

[*]Department of Mathematics, National University of Singapore, Singapore 119076 (matbaowz@nus.edu.sg, URL: http://www.math.nus.edu.sg/~bao/). This author's research was supported by the Ministry of Education of Singapore grant R-146-000-223-112.

[†]Corresponding author. E-mail: jiangwei1007@whu.edu.cn. School of Mathematics and Statistics & Computational Science Hubei Key Laboratory, Wuhan University, Wuhan 430072, P.R. China. This author's research was supported by the National Natural Science Foundation of China Nos. 11401446 and 91630313.

[‡]Departments of Materials Science and Engineering & Mechanical Engineering and Applied Mechanics, University of Pennsylvania, Philadelphia, PA 19104, USA (srol@seas.upenn.edu).

[§]Applied and Computational Mathematics Division, Beijing Computational Science Research Center, Beijing 100193, P.R. China (matwyan@csrc.ac.cn). This author's research was supported by the National Natural Science Foundation of China Nos. 91630207 and U1530401.

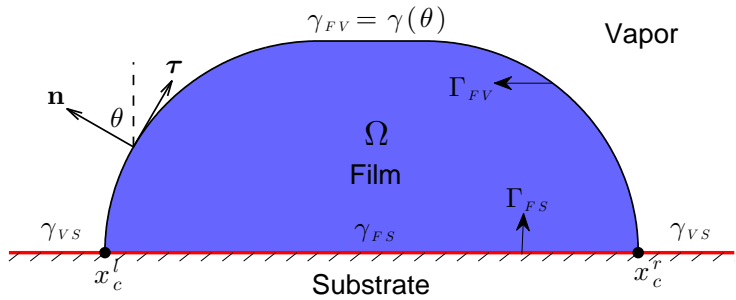

Fig. 1.1. *A schematic illustration of a film or island on a flat, rigid substrate in two dimensions with three interfaces, i.e., film/vapor (FV), film/substrate (FS) and vapor/substrate (VS) interfaces.*

From a mathematical perspective, theoretical solid-state dewetting studies can be categorized into two major classes. The first focuses on the equilibrium of particles on substrates (e.g. [22, 49]), *i.e.*, finding the stable equilibrium shapes of solid-state particles via constrained minimization; the second examines the temporal evolution of the film/particle morphology (e.g. [5, 15, 16, 37, 41, 44]), *i.e.*, examining dewetting dynamics subject to different classes of kinetic phenomena, such as Rayleigh-like instabilities [15, 18], corner-induced instabilities [50] and periodic mass-shedding [41, 44]. In this paper, we mainly focus on the determination of equilibrium shapes of anisotropic particles on substrates and validate the theoretical predictions on (multiple) stable equilibria by numerically solving sharp-interface dynamical evolution models based on surface diffusion and contact line migration.

Following the thermodynamic principles laid out by Gibbs [9], we seek to determine the equilibrium shape of an island/film on a substrate that is a connected shape $\Omega$ which minimizes the total interfacial energy functional of the system [16, 22],

$$(1.1) \quad \min_{\Omega} \ W = \int_{\Gamma_{FV}} \gamma_{FV} d\Gamma_{FV} + \underbrace{\int_{\Gamma_{FS}} \gamma_{FS} d\Gamma_{FS} + \int_{\Gamma_{VS}} \gamma_{VS} d\Gamma_{VS}}_{\text{Substrate Energy}} \quad \text{s.t.} \ |\Omega| = \text{const.}$$

Here, $\Omega$ refers to the island/film (see Fig. 1.1), $|\Omega|$ represents its total volume (3D)/area (2D) of the region, $\Gamma_{FV}$, $\Gamma_{FS}$ and $\Gamma_{VS}$ represent the film/vapor, film/substrate and vapor/substrate interfaces, respectively, and $\gamma_{FV}, \gamma_{FS}, \gamma_{VS}$ represent their corresponding interfacial energy densities. As is typical in solid-state dewetting problems, we assume that $\gamma_{FS}$ and $\gamma_{VS}$ are constants (we assume here that the substrate is flat and remains so during the entire dewetting process, but the effects of a mechanically or chemically reactive substrate can be also included in the model). If $\gamma_{FV}$ is a constant, the problem is isotropic; if $\gamma_{FV}$ depends on the orientation of the film/vapor interface (free surface), it is anisotropic. For crystalline films, $\gamma_{FV}$ is a function of the surface normal (the angle between the outer surface normal and the $y$-axis in 2D) $\theta$ of the film/vapor interface, i.e., $\gamma_{FV} := \gamma(\theta) \in C^0[-\pi, \pi]$ is a positive periodic function. For $\gamma(\theta) \in C^2[-\pi,\pi]$, the surface energy is referred to as smooth; otherwise, it is non-smooth or "cusped". For a smooth surface energy $\gamma(\theta)$, it is weakly anisotropic if the surface stiffness $\widetilde{\gamma}(\theta) := \gamma(\theta) + \gamma''(\theta) > 0$ for all $\theta \in [-\pi,\pi]$; otherwise, it is strongly anisotropic. We remark here that, in the strongly anisotropic case, one approach in the literature is to "modify" the surface energy $\gamma(\theta)$ such that $\widetilde{\gamma}(\theta)$ is positive for all $\theta \in [-\pi,\pi]$ [6, 7].

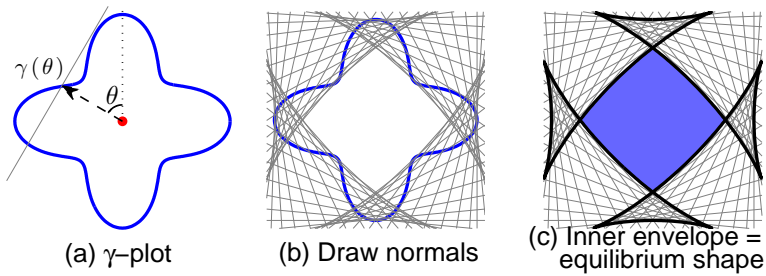

Fig. 1.2. *A schematic illustration for the Wulff construction in 2D with a strongly anisotropic surface energy $\gamma(\theta) = 1 + 0.3\cos(4\theta)$, where (a) is the $\gamma$-plot (blue curve of points with angle $\theta$ and distance $\gamma(\theta)$ to the origin, i.e., the Wulff point) with the gray line (normal) perpendicular to the radius vector; (b) shows these normals at all points along the $\gamma$-plot; (c) shows the equilibrium particle (Wulff) shape in blue (i.e., the solution to (1.2)) - this is the inner envelope of all of these planes (outlined in black) after the "ears" have been removed.*

In the absence of a substrate, the island/film is completely surrounded by the vapor and the problem reduces to the classical problem in applied mathematics and materials science, i.e., determining a connected shape $\Omega$ such that

$$(1.2) \qquad \min_{\Omega} \ W = \int_{\Gamma_{FV}} \gamma_{FV} d\Gamma_{FV} \quad \text{for} \quad |\Omega| = \text{const.}$$

This problem was first solved by Wulff [45] over a century ago using a geometrical approach which has become known as the Wulff (or Gibbs-Wulff) construction (see an illustration in Fig. 1.2). In fact, for the isotropic and weakly anisotropic cases, the Wulff envelope yields a unique connected particle shape which is the equilibrium; however, for the strongly anisotropic case, the Wulff envelope includes distinct "ears", and its inner simply-connected envelope yields the equilibrium by applying the van Gogh treatment, i.e., cutting off/removing these "ears" (cf. Fig. 1.2(c)). For the convenience of readers, Fig. 1.3 displays the Wulff constructions for several different types of surface energy densities $\gamma(\theta)$.

Subsequently, several researchers (e.g., see [8, 25, 27, 31, 38]) provided rigorous proofs that the Wulff construction indeed yields a particle shape of minimal energy, i.e. the solution of (1.2). For example, Fonseca and Muller [8] applied geometric measure theory to show that the Wulff construction gives the equilibrium (solution) to (1.2). On the other hand, Gurski, McFadden and Miksis determined the linear stability of the above problem under the weak anisotropy, and they considered first and second variations of a two-dimensional equilibrium shape, while including the important effects of an axial perturbation (i.e., Rayleigh instability) [10, 11]. In fact, the variational problem (1.2) is scale-invariant and thus it can be first solved for any fixed volume/area in 3D/2D and followed by an appropriate re-scaling. In 2D, if

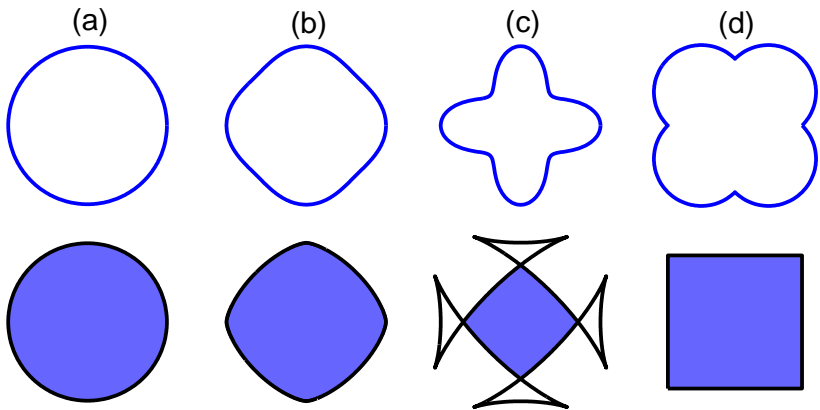

FIG. 1.3. *γ-plot (blue line), Wulff envelope (black line) and Wulff shape (shaded blue region) for several typical surface energy densities $\gamma(\theta)$: (a) $\gamma(\theta) \equiv 1$ (isotropic); (b) $\gamma(\theta) = 1 + 0.06\cos(4\theta)$ (weakly anisotropic); (c) $\gamma(\theta) = 1 + 0.3\cos(4\theta)$ (strongly anisotropic); and (d) $\gamma(\theta) = |\cos\theta| + |\sin\theta|$ ("cusped").*

$\gamma(\theta) \in C^1[-\pi, \pi]$, the Wulff envelope can be expressed analytically as [14, 32, 41]:

$$(1.3) \qquad \begin{cases} x(\theta) = -\gamma(\theta)\sin\theta - \gamma'(\theta)\cos\theta, \\ y(\theta) = \gamma(\theta)\cos\theta - \gamma'(\theta)\sin\theta, \end{cases} \qquad \theta \in [-\pi, \pi],$$

where a scaling factor can be chosen such that the area of the inner simply-connected region surrounded by the curve is $|\Omega|$.

The equilibrium shape of an island on a substrate (*i.e.*, the island in contact with a flat substrate where the substrate energy terms are included) is the solution to (1.1) and can be determined using the geometrical Wulff-Kaischew construction [19]. This equilibrium shape is often called the Winterbottom shape [43], *i.e.*, a Wulff shape truncated by a flat substrate plane/line in 3D/2D, and where the Wulff shape is truncated depends on the wettability of the substrate, *i.e.*, $\gamma_{VS} - \gamma_{FS}$. The Winterbottom construction, though widely studied and applied (e.g., [21, 23, 48]), only describes island shapes corresponding to the global minimum of the free energy, *i.e.* the solution to (1.1). It does not, however, predict all possible stable equilibria, including metastable solutions - local minimizers. In fact, experiments often show island shapes that are not consistent with the Winterbottom shape and this discrepancy is likely attributable to the island shapes trapped in metastable free energy minima [24, 28]. To clarify this issue, this paper addresses the determination of all possible stable equilibria (including local and global minimizers) of the problem (1.1). We propose a generalization of the Winterbottom construction that is not only simply the truncated Wulff shape, but also considers the "ears" from the truncated Wulff envelope.

This paper is organized as follows. In section 2, we consider first-order perturbations to the shape of the two-dimensional film/vapor interface and contact positions, including the first and second thermodynamic variations of the total interfacial energy functional. Based on the thermodynamic variations, in section 3 we propose a generalized Winterbottom construction for determining all possible stable equilibrium shapes. In section 4, we adopt a sharp-interface dynamical evolution model under surface diffusion and contact line migration for simulating solid-state dewetting and implement this model by a numerical method from different initial conditions for find-ing stable equilibrium shapes of the problem. Then, several numerical simulations are reported to validate the generalized Winterbottom construction in section 5. Finally, we draw some conclusions in section 6.

**2. Thermodynamic variation.** In this section, we consider the solid-state dewetting problem in 2D and calculate its first and second thermodynamic variations of the total free energy of the system. For simplicity, all the physical variables in the following discussion are non-dimensionalized, and the subscript $s$ stands for differentiation with respect to the arc length.

As illustrated in Fig. 2.1, the film/vapor interface is denoted as $\Gamma := \mathbf{X}(s) = (x(s), y(s))$, $s \in [0, L]$ with arc length $s$, and $L$ represents the total length of the interface $\Gamma$. The outer unit normal vector $\mathbf{n}$ and unit tangent vector $\boldsymbol{\tau}$ can be expressed as: $\mathbf{n} = (-y_s, x_s)$ and $\boldsymbol{\tau} = (x_s, y_s)$. Suppose that the flat substrate coincides with the $x$-axis of a Cartesian system, and $x_c^l$ and $x_c^r$ represent the $x$-axis coordinates of the left and right contact points, respectively, *i.e.*, $x(0) = x_c^l$ and $x(L) = x_c^r$. In this scenario, the constrained minimization problem (1.1) can be reformulated as (by subtracting a constant and in dimensionless form):

$$(2.1) \qquad \min_{\Omega} \ W := W(\Gamma) = \int_{\Gamma} \gamma(\theta) \, d\Gamma - \sigma(x_c^r - x_c^l) \quad \text{s.t.} \quad |\Omega| = \text{const},$$

where $\sigma := (\gamma_{VS} - \gamma_{FS})/\gamma_0$ is a dimensionless material constant, $\gamma_0$ is a surface energy density employed to non-dimensionalize the equations, and $\gamma(\theta)$ is the dimensionless film/vapor surface energy density (scaled by $\gamma_0$). In the following discussion, according to (1.3), we assume that $\gamma(\theta) \in C^3([-\pi, \pi])$ so that the equilibrium curve is at least a continuous piecewise-$C^2$ curve.

**2.1. First and second variations.** We calculate the first and second variations of the energy functional $W$ defined in Eq. (2.1) with respect to the interface curve $\Gamma$ and the left and right contact points, $x_c^l$ and $x_c^r$. In general, the calculation of the first variation with respect to a closed curve, such as the functional defined in the problem (1.2) [4] and the Willmore functional [42], only requires consideration of the perturbation along the normal direction of the closed curve, because the perturbation along the tangent direction of the closed curve contributes nothing to the first variation. However, in this case, the energy functional also depends on the two contact points ($x_c^l$ and $x_c^r$), and we find that the tangent perturbation plays an important role in understanding contact point perturbations; hence, tangent perturbation must also be included.

We assume that the interface $\Gamma = (x(s), y(s))$ is a continuous piecewise-$C^2$ curve, and its derivatives about the arc length, in the following discussion, can be understood in the weak sense. We consider the following infinitesimal first-order perturbation about the curve $\Gamma$ along both its normal and tangent directions (see Fig. 2.1):

$$(2.2) \qquad \Gamma^\epsilon \ = \ \Gamma + \epsilon\varphi(s)\mathbf{n} + \epsilon\psi(s)\boldsymbol{\tau},$$

where $0 < \epsilon \ll 1$ is a small perturbation parameter, $\varphi(s), \psi(s) \in \text{Lip}[0, L]$ represent Lipschitz continuous perturbation functions with respect to arc length $s$. Then the two components of the new curve $\Gamma^\epsilon$ can be expressed as follows:

$$(2.3) \qquad \Gamma^\epsilon = (x^\epsilon(s), \ y^\epsilon(s)) = (x(s) + \epsilon u(s), \ y(s) + \epsilon v(s)), \qquad 0 \le s \le L,$$

where $s$ is still a parameterization (but not the arc length) of the new curve $\Gamma^\epsilon$ and

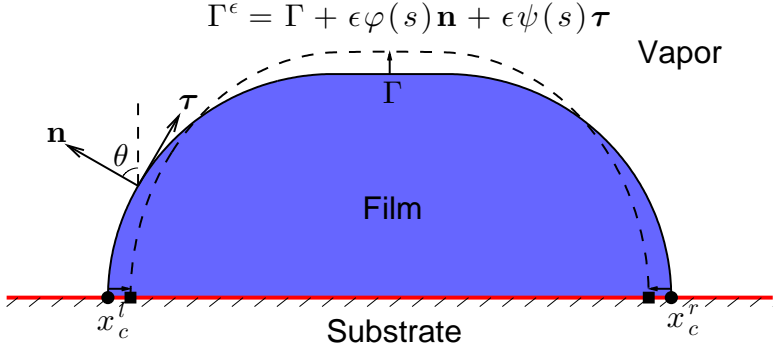

FIG. 2.1. *A schematic illustration of an infinitesimal perturbation (denoted by the dashed line) about the film/vapor interface curve $\Gamma$ along its normal and tangent directions. As the curve is perturbed, the contact points $x_c^l$ and $x_c^r$ may move along the substrate.*

its two component increments along the $x$- and $y$-axis are defined as

(2.4)
$$\begin{cases} u(s) = x_s(s)\psi(s) - y_s(s)\varphi(s), \\ v(s) = x_s(s)\varphi(s) + y_s(s)\psi(s). \end{cases}$$

Equivalently, the functions $\varphi(s)$ and $\psi(s)$ can be expressed by

(2.5)
$$\begin{cases} \varphi(s) = x_s(s)v(s) - y_s(s)u(s), \\ \psi(s) = x_s(s)u(s) + y_s(s)v(s). \end{cases}$$

As illustrated in Fig. 2.1, because the contact points must move along the substrate, the increments along the $y$-axis at the two contact points must be zero, *i.e.*,

(2.6) $\quad v(0) = v(L) = 0 \iff x_s(0)\varphi(0) + y_s(0)\psi(0) = x_s(L)\varphi(L) + y_s(L)\psi(L) = 0.$

Due to this infinitesimal perturbation, the total interface free energy $W(\Gamma^\epsilon) := W(\Gamma^\epsilon; \varphi, \psi)$ for the new curve $\Gamma^\epsilon$ after perturbation becomes:

$$W(\Gamma^\epsilon) := W(\Gamma^\epsilon; \varphi, \psi) = \int_{\Gamma^\epsilon} \gamma(\theta^\epsilon) \, d\Gamma^\epsilon - \sigma\Big[\big(x_c^r + \epsilon u(L)\big) - \big(x_c^l + \epsilon u(0)\big)\Big]$$

(2.7)
$$= \int_0^L \gamma\Big(\arctan \frac{y_s^\epsilon}{x_s^\epsilon}\Big)\sqrt{(x_s^\epsilon)^2 + (y_s^\epsilon)^2}\, ds - \sigma\Big[\big(x_c^r + \epsilon u(L)\big) - \big(x_c^l + \epsilon u(0)\big)\Big],$$

where $y_s^\epsilon = y_s + \epsilon v_s$ and $x_s^\epsilon = x_s + \epsilon u_s$, $\theta^\epsilon := \arctan \frac{y_s^\epsilon}{x_s^\epsilon} \in [-\pi, \pi]$ and the function "arctan" is the generalization of the usual arctangent function [16].

If the total energy $W(\Gamma^\epsilon)$ is understood as a function of the perturbation parameter $\epsilon$, then by simple calculations, we obtain

$$\frac{dW(\Gamma^\epsilon)}{d\epsilon} = \int_0^L \left[\gamma'(\theta^\epsilon)\frac{d\theta^\epsilon}{d\epsilon}\sqrt{(x_s^\epsilon)^2 + (y_s^\epsilon)^2} + \gamma(\theta^\epsilon)\frac{x_s u_s + y_s v_s + \epsilon(u_s^2 + v_s^2)}{\sqrt{(x_s^\epsilon)^2 + (y_s^\epsilon)^2}}\right] ds$$

(2.8)
$$\qquad - \sigma\Big[u(L) - u(0)\Big],$$

where

$$\frac{d\theta^\epsilon}{d\epsilon} = \frac{x_s v_s - y_s u_s}{(x_s^\epsilon)^2 + (y_s^\epsilon)^2}. \tag{2.9}$$

Noting Eqs. (2.4), (2.8) and (2.9), we can calculate the rate of change of the total interface energy functional with respect to the curve $\Gamma(t)$ in terms of $\epsilon$ as

$$\delta W(\Gamma; \varphi, \psi) = \lim_{\epsilon \to 0} \frac{W(\Gamma^\epsilon) - W(\Gamma)}{\epsilon} = \frac{dW(\Gamma^\epsilon)}{d\epsilon}\Big|_{\epsilon=0}$$

$$= \int_0^L \Big[\gamma'(\theta)(\varphi_s - \kappa\psi) + \gamma(\theta)\kappa\varphi + \gamma(\theta)\psi_s\Big] ds - \sigma\Big[u(L) - u(0)\Big]$$

$$= \Big(\gamma'(\theta)\varphi\Big)\Big|_{s=0}^{s=L} + \int_0^L \gamma''(\theta)\kappa\varphi\, ds - \int_0^L \gamma'(\theta)\kappa\psi\, ds + \int_0^L \gamma(\theta)\kappa\varphi\, ds$$

$$+ \Big(\gamma(\theta)\psi\Big)\Big|_{s=0}^{s=L} + \int_0^L \gamma'(\theta)\kappa\psi\, ds - \sigma\Big[u(L) - u(0)\Big]$$

$$= \int_0^L \Big(\gamma''(\theta) + \gamma(\theta)\Big)\kappa\varphi\, ds + \Big[\gamma'(\theta)\varphi(s) + \gamma(\theta)\psi(s) - \sigma u(s)\Big]_{s=0}^{s=L}, \tag{2.10}$$

where $\delta W(\Gamma; \varphi, \psi)$ is called as the first variation of the interface energy functional (2.1) and $\kappa$ represents the curvature of the curve, i.e., $\kappa = -y_{ss}x_s + x_{ss}y_s$.

Assume that $\theta_a^l$ and $\theta_a^r$ are contact angles at the left and right contact points, respectively, then we have

$$x_s(0) = \cos\theta_a^l, \quad y_s(0) = \sin\theta_a^l, \quad x_s(L) = \cos\theta_a^r, \quad y_s(L) = \sin\theta_a^r. \tag{2.11}$$

From Eqs. (2.5) and (2.6), we have

$$\begin{cases} \psi(0) = u(0)\cos\theta_a^l, & \varphi(0) = -u(0)\sin\theta_a^l, \\ \psi(L) = u(L)\cos\theta_a^r, & \varphi(L) = -u(L)\sin\theta_a^r. \end{cases} \tag{2.12}$$

Inserting (2.12) into (2.10), we obtain

$$\delta W(\Gamma; \varphi, \psi) = \int_0^L \mu(s)\varphi(s)\, ds + f(\theta_a^r; \sigma)u(L) - f(\theta_a^l; \sigma)u(0), \tag{2.13}$$

where

$$f(\theta; \sigma) := \gamma(\theta)\cos\theta - \gamma'(\theta)\sin\theta - \sigma, \tag{2.14}$$

and $\mu := \mu(s)$ represents the dimensionless chemical potential of the system,

$$\mu(s) := \widetilde{\gamma}(\theta)\kappa(s) = \big(\gamma''(\theta) + \gamma(\theta)\big)\kappa(s), \tag{2.15}$$

and $\theta := \theta(s)$ is the local orientation of the interface curve.

Similar to the first variation, we calculate the second variation of the energy functional (2.1). From Eqs. (2.8) and (2.9), we have

$$\frac{d^2 W(\Gamma^\epsilon)}{d\epsilon^2} = \int_0^L \bigg\{\gamma''(\theta^\epsilon)\Big(\frac{d\theta^\epsilon}{d\epsilon}\Big)^2 \sqrt{(x_s^\epsilon)^2 + (y_s^\epsilon)^2} + \gamma'(\theta)\frac{d^2\theta^\epsilon}{d\epsilon^2}\sqrt{(x_s^\epsilon)^2 + (y_s^\epsilon)^2}$$

$$+ 2\gamma'(\theta^\epsilon)\frac{d\theta^\epsilon}{d\epsilon}\frac{x_s u_s + y_s v_s + \epsilon(u_s^2 + v_s^2)}{\sqrt{(x_s^\epsilon)^2 + (y_s^\epsilon)^2}} + \gamma(\theta^\epsilon)\frac{u_s^2 + v_s^2}{\sqrt{(x_s^\epsilon)^2 + (y_s^\epsilon)^2}}$$

$$- \gamma(\theta^\epsilon)\frac{\big[x_s u_s + y_s v_s + \epsilon(u_s^2 + v_s^2)\big]^2}{\big[(x_s^\epsilon)^2 + (y_s^\epsilon)^2\big]^{3/2}}\bigg\} ds, \tag{2.16}$$

where $x_s^\epsilon = x_s + \epsilon u_s$, $y_s^\epsilon = y_s + \epsilon v_s$, and $\frac{d\theta^\epsilon}{d\epsilon}$ is defined in Eq. (2.9) and

$$(2.17) \qquad \frac{d^2\theta^\epsilon}{d\epsilon^2} = -2\frac{(x_s v_s - y_s u_s)\left[x_s u_s + y_s v_s + \epsilon(u_s^2 + v_s^2)\right]}{\left[(x_s^\epsilon)^2 + (y_s^\epsilon)^2\right]^2}.$$

Substituting (2.5) and (2.17) into (2.16) and recalling that $\kappa = -y_{ss}x_s + x_{ss}y_s$ and $x_s^2 + y_s^2 = 1$, we obtain

$$\delta^2 W(\Gamma; \varphi, \psi) = \frac{d^2 W(\Gamma^\epsilon)}{d\epsilon^2}\bigg|_{\epsilon=0}$$
$$= \int_0^L \left\{\gamma''(\theta)(x_s v_s - y_s u_s)^2 + \gamma(\theta)\left[(u_s^2 + v_s^2) - (x_s u_s + y_s v_s)^2\right]\right\} ds$$
$$(2.18) \qquad = \int_0^L \left[\gamma(\theta) + \gamma''(\theta)\right](x_s v_s - y_s u_s)^2\, ds = \int_0^L \widetilde{\gamma}(\theta)(\varphi_s - \kappa\psi)^2\, ds.$$

**2.2. Equilibrium shapes and stability.** Based on the above first variation of the total free energy functional, *i.e.*, Eq. (2.13), and by assuming the normal perturbation function $\varphi(s)$ satisfies

$$(2.19) \qquad \int_0^L \varphi(s)\, ds = 0$$

to ensure the total area/mass conservation in the first-order sense, we obtain necessary and sufficient conditions for equilibrium shapes of the solid-state dewetting problem (2.1).

DEFINITION 2.1. (**Equilibrium shapes**): *If a continuous piecewise-$C^2$ curve $\Gamma_e := \big(x(s),\, y(s)\big)$ for $s \in [0, L]$, satisfies $\delta W(\Gamma_e; \varphi, \psi) \equiv 0,\ \forall\ \varphi, \psi \in Lip[0, L]$ with the perturbation function $\varphi(s)$ satisfying Eq. (2.19), i.e., $\int_0^L \varphi(s)\, ds = 0$, it is an equilibrium shape of the solid-state dewetting problem (2.1).*

LEMMA 2.2. *Assume that a continuous piecewise-$C^2$ curve $\Gamma_e := \big(x(s),\, y(s)\big)$ for $s \in [0, L]$, represents the film/vapor interface of the solid-state dewetting problem (2.1) with material constant $\sigma$ and surface energy density $\gamma(\theta)$. Then $\Gamma_e$ is an equilibrium shape if and only if the following two conditions are satisfied:*

$$(2.20) \qquad \mu(s) := \widetilde{\gamma}(\theta)\kappa(s) = [\gamma(\theta) + \gamma''(\theta)]\,\kappa(s) \equiv C, \qquad a.e.\ \ s \in [0, L],$$
$$(2.21) \qquad f(\theta; \sigma) = \gamma(\theta)\cos\theta - \gamma'(\theta)\sin\theta - \sigma = 0, \qquad \theta = \theta_a^l, \theta_a^r,$$

*where the constant $C$ is determined by the given area of the film, and $\theta_a^l \in [0, \pi]$ and $\theta_a^r \in [-\pi, 0]$ represent the left and right contact angles of $\Gamma_e$, respectively.*

*Proof.* It's easy to see that if the conditions (2.20)-(2.21) are both satisfied, by using Eqs. (2.13) and (2.19), then we find that $\delta W(\Gamma_e; \varphi, \psi) \equiv 0,\ \forall \varphi, \psi \in \text{Lip}[0, L]$. Therefore, $\Gamma_e$ is an equilibrium shape.

Conversely, if $\Gamma_e$ is an equilibrium shape, then $\delta W(\Gamma_e; \varphi, \psi) \equiv 0, \forall \varphi, \psi \in \text{Lip}[0, L]$. Since $u(0)$ and $u(L)$ are arbitrary, by using Eq. (2.13), we can obtain the condition (2.21) and the following expression

$$\int_0^L \mu(s)\varphi(s)\, ds = 0.$$

By choosing $C = \frac{1}{L} \int_0^L \mu(s)\,ds$ and making use of $\int_0^L \varphi(s)\,ds = 0$, we have $\int_0^L \big(\mu(s) - C\big)\varphi(s)\,ds = 0$. By choosing $\varphi(s) = \mu(s) - C$, we obtain

$$\int_0^L \big(\mu(s) - C\big)^2\,ds = 0.$$

Therefore, we find that $\mu(s) \equiv C$, a.e. $s \in [0, L]$, i.e., the condition (2.20). □

It can be easily shown that the Wulff envelope, i.e., Eq. (1.3), satisfies the condition (2.20): $\mu(s) \equiv C$. Note that when $\gamma(\theta)$ is strongly anisotropic, the curve given above by Eq. (1.3) will self-intersect, i.e., the Wulff envelope will form "ears". In this case, without considering the substrate energy, the Wulff construction states that cutting off all "ears" will give the unique equilibrium shape. Considering the substrate energy, the (classical) Winterbottom construction still follows the Wulff construction approach, however, in the following we show that this is not always true.

Condition (2.21) is called as the anisotropic Young equation, which has been derived over the years by many authors (e.g., see [13, 29, 41]). Its roots determine the anisotropic Young angles, $\theta_a \in [-\pi, \pi]$. For the isotropic case, i.e., $\gamma(\theta) \equiv 1$, it is easy to see that the anisotropic Young equation collapses to the well-known (isotropic) Young equation, i.e., $\cos\theta = \sigma = \frac{\gamma_{VS} - \gamma_{FS}}{\gamma_0}$, which is commonly used in liquid-state wetting/dewetting [47, 3]. When $-1 \leq \sigma \leq 1$, the Young equation has a unique root $\theta_a^l = \cos^{-1}\sigma$ in the interval $[0, \pi]$ and a unique root $\theta_a^r = -\cos^{-1}\sigma$ in the interval $[-\pi, 0]$; and when $|\sigma| > 1$, there is no root and complete wetting or dewetting occurs. For the anisotropic case, differentiating $f(\theta; \sigma)$ with respect to $\theta$, we obtain

$$\frac{df(\theta; \sigma)}{d\theta} = -[\gamma''(\theta) + \gamma(\theta)]\sin\theta = -\widetilde{\gamma}(\theta)\sin\theta, \quad \theta \in [-\pi, \pi].$$

For the weakly anisotropic case, i.e., the surface stiffness $\widetilde{\gamma}(\theta) = \gamma''(\theta) + \gamma(\theta) > 0$ for $\theta \in [-\pi, \pi]$, $f(\theta; \sigma)$ is a monotonously decreasing function on the interval $[0, \pi]$, and a monotonously increasing function on the interval $[-\pi, 0]$, respectively; in this case, for any given $\sigma \in \mathbb{R}$, the anisotropic Young equation (2.21) has at most one root $\theta = \theta_a^l \in [0, \pi]$, and at most one root $\theta = \theta_a^r \in [-\pi, 0]$, respectively. On the other hand, in the strongly anisotropic case, i.e., the surface stiffness $\widetilde{\gamma}(\theta) = \gamma''(\theta) + \gamma(\theta)$ changes sign over $[-\pi, \pi]$ and the anisotropic Young equation (2.21) may have multiple roots over the interval $[0, \pi]$ (see Fig. 2.2) and/or over the interval $[-\pi, 0]$.

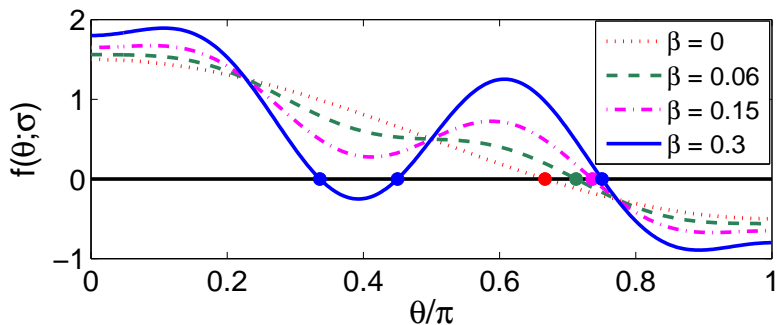

FIG. 2.2. Plot of $f(\theta; \sigma)$ as a function of $\theta$ for $\gamma(\theta) = 1 + \beta\cos(4\theta)$ for different $\beta$ and $\sigma = -0.5$.

Looking back at the explicit expression of the Wulff envelope, i.e., Eq. (1.3), if we first use a flat substrate line given by the expression $y = \sigma$ to truncate the Wulff envelope and then redefine the origin of the Cartesian coordinates for the Wulff envelope by denoting the $y$-coordinate of the flat substrate line as the zero (the $x$-coordinate is unchanged), we find that the substrate intersects the Wulff envelope at contact points $(x(\theta_a^l), 0)$ and $(x(\theta_a^r), 0)$ with contact angles $\theta_a^l \in [0, \pi]$ and $\theta_a^r \in [-\pi, 0]$. It is easy to see that $\theta_a^l$ and $\theta_a^r$ satisfy the anisotropic Young equation (2.21). Therefore, we conclude that a connected curve segment of the Wulff envelope, which is truncated by a flat substrate line $y = \sigma$, is an equilibrium shape of the solid-state dewetting problem (2.1) [16]. However, the situation is much more complicated when the surface energy is strongly anisotropic due to the existence of multiple roots over $[0, \pi]$ and/or $[-\pi, 0]$ in the anisotropic Young equation (2.21). In general, not all equilibria with anisotropic Young contact angles (roots of the anisotropic Young equation (2.21))) are stable. Stability conditions for the equilibrium shapes can be determined by using the second variation of the energy functional.

DEFINITION 2.3. (**Stable equilibrium**): *If a continuous piecewise-$C^2$ curve $\Gamma_e := \bigl(x(s),\ y(s)\bigr)$ for $s \in [0, L]$ satisfies the relation: $\forall\ \nu_0 > 0$, there exists a small positive number $\epsilon_0$, such that when $|\epsilon| < \epsilon_0$, the following relation always holds:*

$$(2.22) \quad W(\Gamma_e) \leq W(\Gamma_e^\epsilon; \varphi, \psi) \leq W(\Gamma_e) + \nu_0, \quad \forall\ \|\varphi\|_{Lip[0,L]} \leq 1,\ \|\psi\|_{Lip[0,L]} \leq 1,$$

*where $W(\Gamma_e^\epsilon; \varphi, \psi)$ is defined above in Eq. (2.7) with the perturbation function $\varphi(s)$ satisfying Eq. (2.19) (i.e., $\int_0^L \varphi(s)\, ds = 0$), then $\Gamma_e$ is a stable equilibrium of the solid-state dewetting problem (2.1).*

According to the above second variation, *i.e.*, Eq. (2.18), we can obtain a necessary condition for stable equilibria of the solid-state dewetting problem (2.1).

LEMMA 2.4. *If a continuous piecewise-$C^2$ curve $\Gamma_e := \bigl(x(s),\ y(s)\bigr)$ for $s \in [0, L]$ satisfies relation (2.22), then $\Gamma_e$ is an equilibrium shape and the following stability condition (with respect to two-dimensional perturbations) holds:*

$$(2.23) \quad \widetilde{\gamma}(\theta(s)) = \gamma(\theta) + \gamma\,''(\theta) \geq 0, \qquad a.e.\ \ s \in [0, L].$$

*Proof.* If $\Gamma_e$ satisfies (2.22), then when $|\epsilon| < \epsilon_0$, $W(\Gamma_e) \leq W(\Gamma_e^\epsilon; \varphi, \psi)$. By examining $W(\Gamma_e^\epsilon; \varphi, \psi)$ as a function of $\epsilon$ and $W(\Gamma_e) = W(\Gamma_e^0; \varphi, \psi)$, we see that

$$\delta W(\Gamma_e; \varphi, \psi) = \lim_{\epsilon \to 0} \frac{1}{\epsilon}\bigl[W(\Gamma_e^\epsilon; \varphi, \psi) - W(\Gamma_e)\bigr] = 0.$$

Therefore, $\Gamma_e$ is an equilibrium shape of the solid-state dewetting problem (2.1).

Furthermore, using Taylor expansion for $W(\Gamma_e^\epsilon; \varphi, \psi)$ with respect to $\epsilon$, we have

$$W(\Gamma_e^\epsilon; \varphi, \psi) = W(\Gamma_e) + \delta W(\Gamma_e; \varphi, \psi)\, \epsilon + \frac{1}{2}\, \delta^2 W(\Gamma_e; \varphi, \psi)\, \epsilon^2 + \mathcal{O}(\epsilon^3),$$

where the second variation $\delta^2 W(\Gamma_e; \varphi, \psi)$ is defined by Eq. (2.18). Then, by using the above definition of stable equilibrium and the first variation $\delta W(\Gamma_e; \varphi, \psi) = 0$, we immediately have

$$\delta^2 W(\Gamma_e; \varphi, \psi) = \int_0^L \widetilde{\gamma}(\theta)\bigl(\varphi_s - \kappa \psi\bigr)^2 ds \geq 0.$$

Since $\varphi(s)$ and $\psi(s)$ are arbitrarily chosen functions, we obtain $\widetilde{\gamma}(\theta(s)) = \gamma(\theta) + \gamma\,''(\theta) \geq 0$, a.e. $s \in [0, L]$. □

REMARK 2.1. *This Lemma shows that all surface orientations presented in the stable equilibrium shape have non-negative surface stiffness. We refer to these orientations as stable orientations.*

Combining Lemmas 2.2 and 2.4, we have the following theorem which presents necessary conditions for stable equilibria of the solid-state dewetting problem (2.1).

THEOREM 2.5. *(**Necessary conditions**): Assume that a continuous piecewise-$C^2$ curve $\Gamma_e := \bigl(x(s),\ y(s)\bigr)$ for $s \in [0, L]$ is the film/vapor interface of the solid-state dewetting problem (2.1) with material constant $\sigma$ and film/vapor interface energy density $\gamma(\theta)$. If $\Gamma_e$ is a stable equilibrium, then the three conditions (2.20), (2.21) and (2.23) are simultaneously satisfied.*

**3. A generalized Winterbottom construction.** In this section, we present a generalization of the Winterbottom construction for obtaining all possible stable equilibria of solid-state dewetting for a given surface energy density $\gamma(\theta)$ and material constant $\sigma$. By applying this generalized Winterbottom construction, we show that for a crystalline material with an $m$-fold symmetric surface energy, there are cases for which multiple stable equilibrium island shapes exist for strong anisotropy [16, 24, 20].

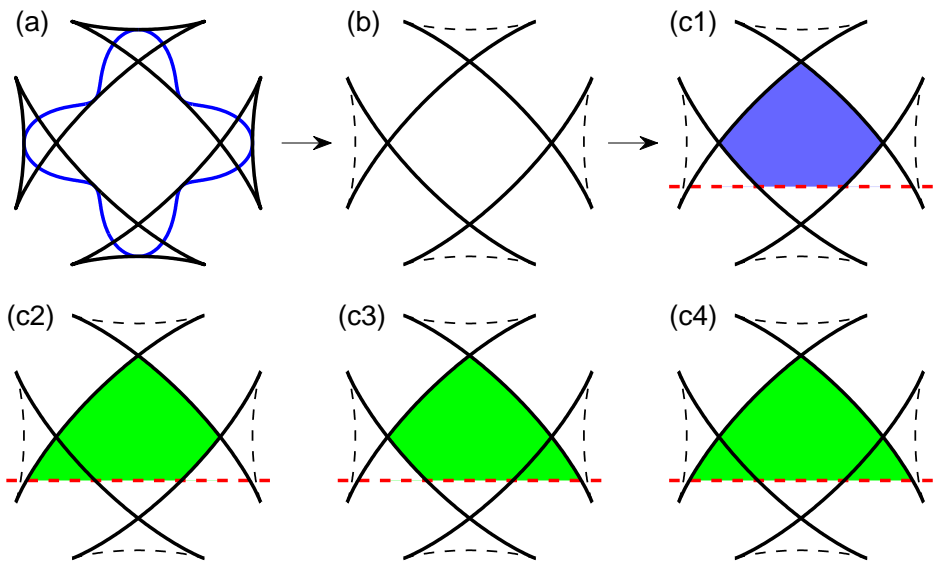

FIG. 3.1. *A simple illustration of the three steps for obtaining all possible stable equilibria: (a) given a $\gamma$-plot (blue solid curves), obtain the corresponding Wulff envelope (black solid curves); (b) then remove all unstable equilibrium orientations (black dash curves); and (c1)-(c4) finally use a flat substrate (red dash line) to truncate the Wulff envelope to obtain stable equilibrium island shapes (blue and green shaded regions). Here, we choose $\gamma(\theta) = 1 + 0.3\cos(4\theta)$ and $\sigma = -0.5$ as an example.*

**3.1. Three steps of the construction.** According to the three conditions presented in Theorem 2.5, if given a surface energy density $\gamma(\theta)$ and a material parameter $\sigma := \dfrac{\gamma_{VS} - \gamma_{FS}}{\gamma_0}$, then we can construct all possible stable equilibrium shapes via the following three simple steps (see Fig. 3.1):

**Step 1.** Condition (2.20) $\implies$ *"Draw the Wulff envelope"*: For a given $\gamma(\theta)$, draw its $\gamma$-plot (blue curve in Fig. 3.1(a)). Obtain its Wulff envelope (black solid curve in Fig. 3.1(a)) from the $\gamma$-plot via the Wulff construction. The Wulff envelope can be obtained analytically from Eqs. (1.3) when $\gamma(\theta)$ is smooth.

**Step 2.** Condition (2.23) $\implies$ *"Remove all unstable orientations"*: Remove all orientations from the Wulff envelope (obtained in *Step 1*) for which the surface stiffness is negative (black dashed curves in Fig. 3.1(b)). As shown in Fig. 3.1(b), only parts of the Wulff envelope "ears" are unstable.

**Step 3.** Condition (2.21) $\implies$ *"Truncate the Wulff envelope"*: Add a flat substrate $y = \sigma$ (red dash line in Fig. 3.1(c1)-(c4)) and discard the sections of the remaining Wulff envelope (obtained in *Step 2*). The origin of the $x$-$y$ coordinates lies at the center of the Wulff envelope (*i.e.*, the Wulff point shown in Fig. 1.2(a)). Stable equilibria are enclosed by the Wulff envelope and the substrate line (blue shaded region in Fig. 3.1(c1) and green shaded regions in Fig. 3.1(c2)-(c4)).

In general, when the surface energy is isotropic or weakly anisotropic (*i.e.*, the Wulff envelope has no "ears"), this procedure can produce at most one stable equilibrium. This is the most widely discussed application of the Winterbottom construction. However, when the surface energy is strongly anisotropic, this procedure may produce multiple stable equilibria, depending on the position of the flat substrate line (*i.e.*, the material constant $\sigma$). For example, if we choose $\sigma = 0.25$, this will produce only one stable equilibrium; on the contrary, if we choose $\sigma = -0.5$ (shown in Fig. 3.1), it will produce four stable equilibria, including two symmetric shapes ((c1) and (c4)) and two asymmetric shapes ((c2) and (c3)). In the following discussion, we focus largely on cases of strong anisotropy.

**3.2. Stable equilibria in strongly anisotropic cases.** Here we discuss more details regarding stable equilibrium shapes found by application of the generalized Winterbottom construction for solid thin films with $m$-fold ($m = 2, 3, 4, 6$) smooth, symmetric film/vapor interface energy densities, which are widely discussed in the materials science literature. We write this surface energy density in the form:

$$(3.1) \qquad \gamma(\theta) = 1 + \beta \cos\left[m(\theta + \phi)\right], \qquad \theta \in [-\pi, \pi],$$

where $\phi \in [0, \pi]$ represents a phase shift angle describing the rotation of the island crystal structure with respect to the substrate plane, and $\beta \ (\geq 0)$ controls the degree of the anisotropy. For this surface energy, when $\beta = 0$ the system is isotropic; when $0 < \beta < \frac{1}{m^2-1}$ it is weakly anisotropic; otherwise, it is strongly anisotropic.

We now focus on four strongly anisotropic cases where the substrate line intersects with part of the Wulff envelope "ears"; these are cases not addressed in the classical Winterbottom construction [19, 43].

**Case** 1 (*Unique stable equilibrium with an inverted shape*). When the flat substrate line intersects with the top "ear" of the Wulff envelope (see Fig. 3.2(a) - (d)), the stable equilibrium can be obtained by "flipping over" the truncated part of the Wulff envelope, as shown in Figs. 3.2(c) and (d). Two different equilibrium island shapes are seen: the first between the substrate line (red) and dashed black line (shaded in green in Fig. 3.2(c)) and the second between the substrate line and the solid black line (shaded in blue in Fig. 3.2(d)). The first one (in green) is unstable, while the second (in blue) is stable. Interestingly, the stable case shown by Fig. 3.2(d) has concave rather than convex surfaces. This is quite different from the classical Winterbottom/Wulff constructions where the island/particle shape is always convex. The convexity of Wulff shapes has been rigorously proven by [31]. In this example, the anisotropic Young equation (2.21) has two different roots (*i.e.*, $\theta_a = \{0.3425, 0.6863\}$) in $[0, \pi]$, but only the root 0.6863 corresponds to the left contact angle of the stable equilibrium shape shown by Fig. 3.2(d).

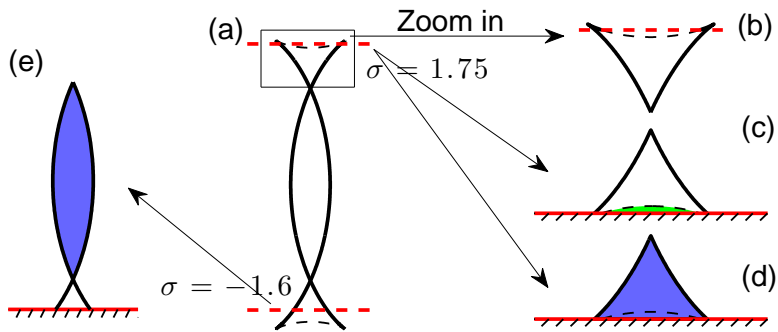

FIG. 3.2. *Generalized Winterbottom constructions for $m = 2, \beta = 0.7, \phi = 0$ with (b)-(d) $\sigma = 1.75$ and (e) $\sigma = -1.6$, where (a) the solid and dashed black curves show the stable equilibrium and unstable equilibrium sections of the Wulff envelope (with its "ears"), and the red dashed lines correspond to the substrate for two different value of $\sigma$; (b) an enlarged section of the upper portion of the plot in (a) around the substrate line $\sigma = 1.75$; (c) an inverted form of (b) - the green shading represents the unstable equilibrium island shape; (d) an inverted form of (b) - the blue shading represents the stable equilibrium island shape; (e) the truncated Wulff envelope shape when $\sigma = -1.6$ - in this case, the two contact points meet each other, and a complete dewetting will occur (i.e., the island becomes a free particle not attached to the substrate).*

**Case** 2 (*A self-intersection curve with complete dewetting*). When $\sigma = -1.6$, the flat substrate line intersects the bottom "ear" (see Fig. 3.2(a) and (e)), and the truncated Wulff envelope shape is a self-intersection curve, shown by Fig. 3.2(e). In this case, the two contact points meet each other, and complete dewetting occurs - this implies that the island becomes an isolate particle, detached from the substrate. Note here that in this example, the anisotropic Young equation (2.21) has only one root (*i.e.*, $\theta_a = 2.2497$), and it is less than $\pi$.

**Case** 3 (*Unique stable equilibrium*). When the flat substrate line intersects the middle "ears" of the Wulff envelope above the center, only one stable equilibrium shape exists and that is the one represented by the classical Winterbottom construction. In this example (*i.e.*, $\gamma(\theta) = 1 + 0.3\cos(4\theta)$ and $\sigma = 0.25$), the anisotropic Young equation (2.21) has three different roots ($\theta_a = \{0.8471, 1.6436, 2.1649\}$) in $[0, \pi]$, but only the root $\theta_a = 0.8471$ corresponds to the left contact angle of the stable equilibrium shape.

**Case** 4 (*Multiple stable equilibria*). In this case, the substrate line intersects the middle "ears" below the center. As shown in Fig. 3.3, it admits three solutions to the anisotropic Young equation (2.21). For the case shown in Fig. 3.3(a) ($m = 4, \beta = 0.3, \phi = 0, \sigma = -0.5$, for which stable equilibrium shapes have been clearly shown by Fig. 3.1), the three solutions for the left contact angle in $[0, \pi]$ are $\theta_a = \{1.0563, 1.4146, 2.3575\}$ two of which are stable. This yields the four stable equilibrium shapes - two symmetric (blue shaded region and striped region in Fig. 3.3(a)) and two asymmetric stable equilibrium shapes (blue shaded region plus the striped regions to the left or right of it in Fig. 3.3(a)). Similarly, for the case shown in Fig. 3.3(b) ($m = 4, \beta = 0.4, \phi = 0, \sigma = -\sqrt{3}/2$), there are three solutions in $[0, \pi]$ for the left contact angle $\theta_a = \{1.0672, 1.3723, 2.4520\}$ two of which are stable. However, this yields one symmetric (striped region in Fig. 3.3(b)) and two asymmetric stable equilibrium shapes (blue shaded region plus the striped regions on the left or right of the center of the blue region in Fig. 3.3(b), together with a completely dewetted particle (the blue region detached from the substrate in Fig. 3.3(b)).

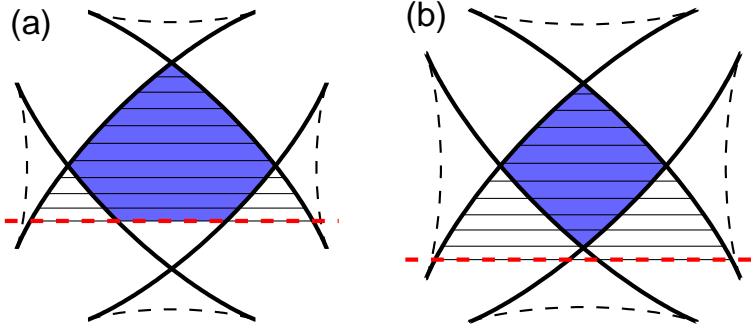

FIG. 3.3. *Two different examples for Case 4, where the generalized Winterbottom constructions are given for (a)* $m = 4, \beta = 0.3, \phi = 0, \sigma = -0.5$; *(b)* $m = 4, \beta = 0.4, \phi = 0, \sigma = -\sqrt{3}/2$.

The four cases discussed here clearly show that the anisotropic Young equation (2.21) may have multiple roots in $[0, \pi]$ (corresponding to static values of the left contact angles), but it is not necessarily the case that all of these are stable; i.e., there are stable and unstable anisotropic Young angles. More precisely, "stable anisotropic Young angles" refer here to all roots to (2.21) for which (2.23) is satisfied, i.e., equilibrium contact angles which may occur for stable orientations. For the $m$-fold crystalline surface energy (3.1) with $\phi = 0$, we find that a stable anisotropic Young angle uniquely determines a symmetric stable equilibrium shape. Hence, the number of stable anisotropic Young angles reflects the number of stable equilibrium island shapes. For a fixed $\phi$, we can examine how the number of anisotropic Young angles (*i.e.*, the roots of the anisotropic Young equation (2.21) in $[0, \pi]$) and the number of those that are stable vary with the magnitude of the anisotropy $\beta$ and the material constant $\sigma$. We show these numbers of anisotropic Young angles as a function of $\beta$ and $\sigma$ for $m$-fold symmetric crystals ($m = 2, 3, 4, 6$) in Fig. 3.4 (for $\phi = 0$). As shown in Fig. 3.4, when $m$ increases, the maximum number of stable anisotropic Young angles also increases, and this means that more stable equilibrium shapes may appear (note: we only show data for $0 \leq \beta < 1$ since outside this range, the surface energy density $\gamma(\theta)$ may be non-positive).

**4. Dynamical evolution via surface diffusion.** In the previous section, we showed that when the surface energy is strongly anisotropic, multiple stable equilibrium island shapes (and contact angles) are possible. Which equilibrium shape appears may depend on how the system evolves, *i.e.,* it may be possible for the system to be trapped in a metastable equilibrium. In order to investigate this question, we propose the dynamical evolution sharp-interface model via surface diffusion-controlled shape evolution for simulating solid-state dewetting and finding numerical stationary (or equilibrium) shapes. In the following section, we present numerical results for this model, evolving the structure until the shape becomes stationary and compare the resulting morphologies with those predicted in the previous section. We previously outline this approach in [41, 16] and describe it here for (i) the isotropic/weakly anisotropic, (ii) the strongly anisotropic, and (iii) the non-smooth and/or "cusped"

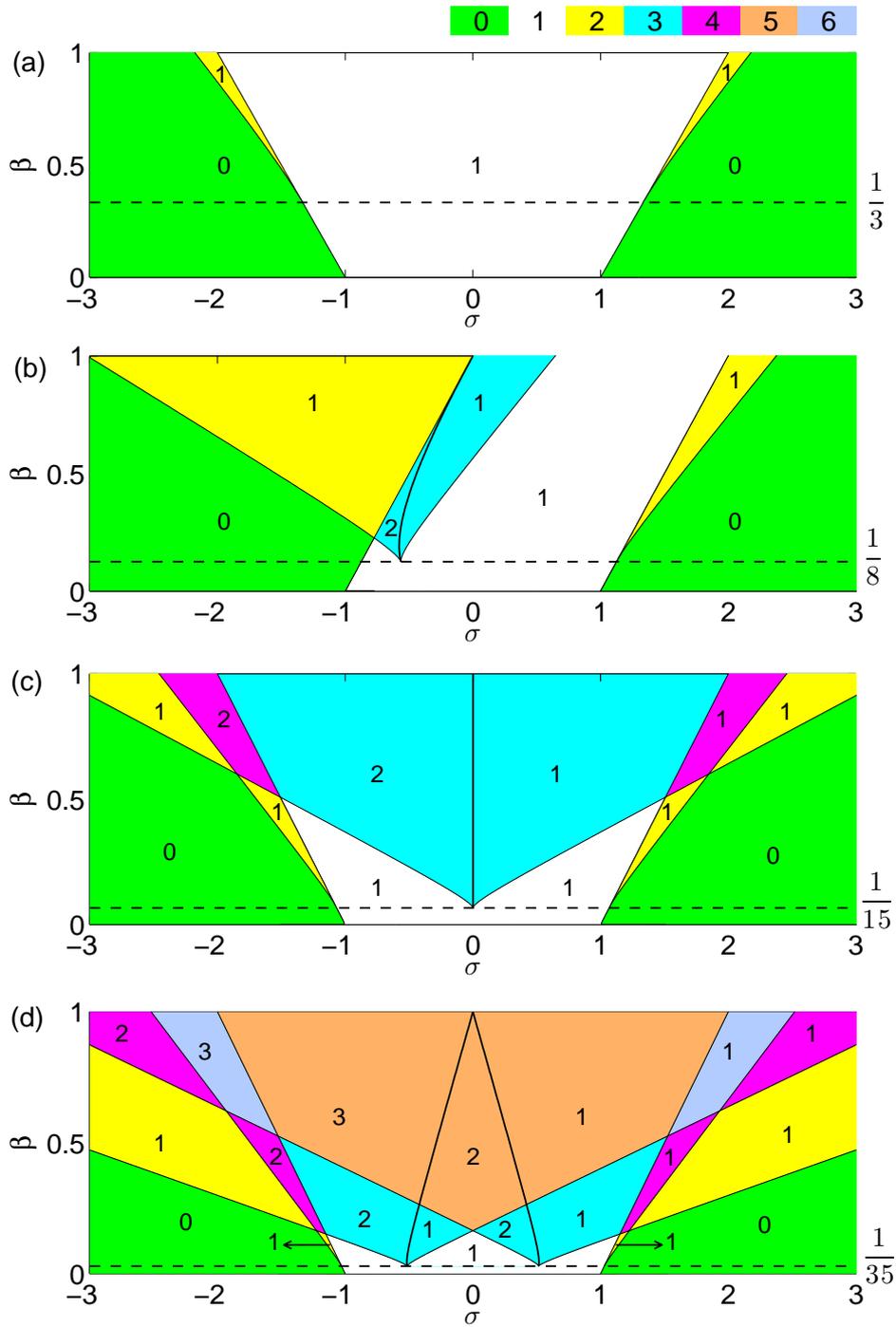

Fig. 3.4. *Phase diagrams of the number of anisotropic Young angles (roots of the anisotropic Young equation (2.21) in $[0, \pi]$) as a function of $\beta$ and $\sigma$ for (a) $m = 2$, (b) $m = 3$, (c) $m = 4$, and (d) $m = 6$ (under the fixed $\phi = 0$). The colors indicate the total number of anisotropic Young angles - see the color bar at the top. The Arabic numerals in each of the plots indicate the number of <u>stable</u> anisotropic Young angles. Of course, the number of <u>stable</u> anisotropic Young angles is less than or equal the number of roots of (2.21).*

surface energy cases.

**4.1. Isotropic/weakly anisotropic case.** In this case, $\gamma(\theta) \in C^2[-\pi, \pi]$ and $\widetilde{\gamma}(\theta) := \gamma(\theta) + \gamma''(\theta) > 0$ for all $\theta \in [-\pi, \pi]$, where the first variation was shown in (2.13). A sharp-interface model for the two-dimensional solid-state dewetting of an island/film on a flat rigid substrate via surface diffusion was proposed in [41, 2]. In dimensionless form, the model may be written as

$$\mathbf{X}_t = \mu_{ss}\, \mathbf{n}, \qquad 0 < s < L(t), \qquad t > 0, \tag{4.1}$$
$$\mu = \left[\gamma(\theta) + \gamma''(\theta)\right]\kappa, \qquad \kappa = -y_{ss}x_s + x_{ss}y_s; \tag{4.2}$$

where $\Gamma := \Gamma(t) = \mathbf{X}(s,t) = (x(s,t), y(s,t))$ represents the moving film/vapor interface, $s$ is the arc length or distance along the interface, $t$ is time, $\mathbf{n} = (-y_s, x_s)$ is the interface outer unit normal direction, $\mu := \mu(s,t)$ is the chemical potential and $L := L(t)$ represents the total length of the moving film/vapor interface. The initial condition is

$$\mathbf{X}(s,0) = (x(s,0), y(s,0)) = \mathbf{X}_0(s) = (x_0(s), y_0(s)), \qquad 0 \le s \le L_0 := L(0). \tag{4.3}$$

and the boundary conditions are [41]:

(i) Contact point condition

$$y(0,t) = 0, \qquad y(L,t) = 0, \qquad t \ge 0; \tag{4.4}$$

(ii) Relaxed contact angle condition

$$\frac{dx_c^l}{dt} = \eta\, f(\theta_d^l; \sigma), \qquad \frac{dx_c^r}{dt} = -\eta\, f(\theta_d^r; \sigma), \qquad t \ge 0, \tag{4.5}$$

where $\theta_d^l := \theta_d^l(t)$ and $\theta_d^r := \theta_d^r(t)$ are the (dynamic) contact angles at the left and right contact points, respectively, $0 < \eta < \infty$ denotes the contact line mobility, and $f(\theta; \sigma)$ is defined as Eq. (2.14);

(iii) Zero-mass flux condition

$$\mu_s(0,t) = 0, \qquad \mu_s(L,t) = 0, \qquad t \ge 0. \tag{4.6}$$

Condition (4.4) implies that the two contact points must move along the flat substrate, condition (4.5) is the energy dissipation relation for the moving contact points [41], and condition (4.6) ensures that the total area/mass of the thin film is conserved (no mass flux at the moving contact points). The total area/mass of the island $A(t)$ and the total interfacial energy $W(t)$ are defined as

$$A(t) = \int_0^{L(t)} y(s,t)\, x_s(s,t)\, ds, \qquad W(t) = \int_0^{L(t)} \gamma(\theta(s,t))\, ds - \sigma[x_c^r(t) - x_c^l(t)]. \tag{4.7}$$

PROPOSITION 4.1. *For the sharp-interface model governing by Eqs.* (4.1)-(4.2) *together with boundary conditions* (4.4)-(4.6) *and the initial condition* (4.3), *the total area/mass is conserved and the total interfacial energy decreases during the evolution in the isotropic or weakly anisotropic case:*

$$A(t) \equiv A(0) = \int_0^{L(0)} y(s,0)\, x_s(s,0)\, ds, \quad t \ge 0, \tag{4.8}$$

$$W(t) \le W(t_1) \le W(0) = \int_0^{L(0)} \gamma(\theta(s,0))\, ds - \sigma[x_c^r(0) - x_c^l(0)], \quad t \ge t_1 \ge 0. \tag{4.9}$$

*Proof.* We first introduce a new time-independent variable $p \in [0,1]$ and reparameterize the moving film/vapor interface such that $p = 0$ and $p = 1$ represent the left and right contact points, respectively. Then the arc length $s$ is a function of $p$ and $t$ and we denote $s_p = \frac{\partial s}{\partial p}$. The area/mass of the thin film in (4.8) can be rewritten as

$$(4.10) \qquad A(t) = \int_0^1 y x_p \, dp.$$

By differentiating (4.10) with respect to $t$ and integrating by parts, we obtain

$$\dot{A}(t) = \int_0^1 (y_t x_p + y x_{pt}) \, dp = \int_0^1 (y_t x_p - y_p x_t) \, dp + y x_t \Big|_{p=0}^{p=1}$$

$$= \int_0^1 (x_t, y_t) \cdot (-y_p, x_p) \, dp = \int_0^{L(t)} \mathbf{X}_t \cdot \mathbf{n} \, ds$$

$$(4.11) \qquad = \int_0^{L(t)} \mu_{ss} \, ds = \mu_s(L(t), t) - \mu_s(0, t) = 0,$$

which immediately implies the total area/mass is conserved.

The total free energy in (4.9) can be rewritten as

$$(4.12) \qquad W(t) = \int_0^1 \gamma(\theta) s_p \, dp - \sigma [x_c^r(t) - x_c^l(t)].$$

Notice that the following equations hold:

$$(4.13) \quad \mathbf{n}_p = \kappa s_p \boldsymbol{\tau}, \quad \boldsymbol{\tau}_p = -\kappa s_p \mathbf{n}, \quad \theta_p = -\kappa s_p, \quad s_{pt} = \mathbf{X}_{pt} \cdot \boldsymbol{\tau}, \quad \theta_t s_p = \mathbf{X}_{pt} \cdot \mathbf{n}.$$

Differentiating (4.12) with respect to $t$ and integrating by parts, making use of the above identities (4.13) and Eqs. (4.1)-(4.6), we obtain

$$\dot{W}(t) = \int_0^1 \left( \gamma'(\theta) \theta_t s_p + \gamma(\theta) s_{pt} \right) dp - \sigma \left( \frac{dx_c^r}{dt} - \frac{dx_c^l}{dt} \right)$$

$$= \int_0^1 \mathbf{X}_{pt} \cdot \left( \gamma'(\theta) \mathbf{n} + \gamma(\theta) \boldsymbol{\tau} \right) dp - \sigma \left( \frac{dx_c^r}{dt} - \frac{dx_c^l}{dt} \right)$$

$$= -\int_0^1 \mathbf{X}_t \cdot \left( \left( \gamma''(\theta) \theta_p \mathbf{n} + \gamma'(\theta) \kappa s_p \boldsymbol{\tau} \right) + \left( \gamma'(\theta) \theta_p \boldsymbol{\tau} - \gamma(\theta) \kappa s_p \mathbf{n} \right) \right) dp$$

$$+ \left[ \mathbf{X}_t \cdot \left( \gamma'(\theta) \mathbf{n} + \gamma(\theta) \boldsymbol{\tau} \right) \right]_{p=0}^{p=1} - \sigma \left( \frac{dx_c^r}{dt} - \frac{dx_c^l}{dt} \right)$$

$$(4.14) \qquad = \int_0^{L(t)} \kappa \bigl( \gamma(\theta) + \gamma''(\theta) \bigr) \mathbf{X}_t \cdot \mathbf{n} \, ds + f(\theta_d^r; \sigma) \frac{dx_c^r}{dt} - f(\theta_d^l; \sigma) \frac{dx_c^l}{dt}$$

$$= \int_0^{L(t)} \mu \mu_{ss} \, ds - \frac{1}{\eta} \left[ \left( \frac{dx_c^r}{dt} \right)^2 + \left( \frac{dx_c^l}{dt} \right)^2 \right]$$

$$= \mu \mu_s \Big|_{s=0}^{s=L(t)} - \int_0^{L(t)} \mu_s^2 \, ds - \frac{1}{\eta} \left[ \left( \frac{dx_c^r}{dt} \right)^2 + \left( \frac{dx_c^l}{dt} \right)^2 \right]$$

$$(4.15) \qquad = -\int_0^{L(t)} \mu_s^2 \, ds - \frac{1}{\eta} \left[ \left( \frac{dx_c^r}{dt} \right)^2 + \left( \frac{dx_c^l}{dt} \right)^2 \right] \leq 0, \qquad t \geq 0,$$

which demonstrates that the total free energy is dissipative during the evolution. □

**4.2. Strongly anisotropic case.** In this case, $\gamma(\theta) \in C^2[-\pi, \pi]$ and the surface stiffness $\widetilde{\gamma}(\theta) := \gamma(\theta) + \gamma''(\theta)$ changes sign for some $\theta \in [-\pi, \pi]$, and the governing equations (4.1)-(4.2) become ill-posed. These governing equations can be regularized by adding regularization terms such that the regularized sharp interface model is well-posed. In practice, this is often done by regularizing the total interfacial energy $W(\Gamma)$ in (2.1) by adding the well-known Willmore energy, i.e., $W_{\text{wm}} = \frac{1}{2} \int_\Gamma \kappa^2 \, d\Gamma$ [12, 40, 16, 2] such that the regularized total interfacial energy becomes

$$(4.16) \qquad W^\varepsilon_{\text{reg}}(\Gamma) = W(\Gamma) + \varepsilon^2 W_{\text{wm}}(\Gamma) = \int_\Gamma \left[ \gamma(\theta) + \frac{\varepsilon^2}{2} \kappa^2 \right] d\Gamma - \sigma(x_c^r - x_c^l),$$

where $0 < \varepsilon \ll 1$ is a small regularization parameter.

Similar to the calculation of the first variation of $W(\Gamma)$ in (2.1) to obtain (2.13), we obtain the first variation of the regularized total interfacial energy $W^\varepsilon_{\text{reg}}(\Gamma)$ as (details omitted here for brevity) [16]:

$$\delta W^\varepsilon_{\text{reg}}(\Gamma; \varphi, \psi) = \int_0^L \mu^\varepsilon(s) \varphi(s) \, ds + f^\varepsilon(\theta_d^r; \sigma) u(L) - f^\varepsilon(\theta_d^l; \sigma) u(0)$$
$$(4.17) \qquad\qquad - (\kappa \varphi_s)|_0^L + \left( \frac{1}{2} \kappa^2 \psi \right) \bigg|_0^L,$$

where the function $f^\varepsilon(\theta; \sigma)$ is defined as

$$(4.18) \qquad f^\varepsilon(\theta; \sigma) := \gamma(\theta) \cos\theta - \gamma'(\theta) \sin\theta - \sigma - \varepsilon^2 \kappa_s(\theta) \sin\theta,$$

and $\mu^\varepsilon := \mu^\varepsilon(s)$ represents the dimensionless (regularized) chemical potential of the system, i.e.,

$$(4.19) \quad \mu^\varepsilon(s) := \widetilde{\gamma}(\theta) \kappa - \varepsilon^2 \left( \frac{\kappa^3}{2} + \kappa_{ss} \right), \quad \widetilde{\gamma}(\theta) = \gamma(\theta) + \gamma''(\theta), \quad \kappa = -y_{ss} x_s + x_{ss} y_s.$$

For $\varepsilon \to 0$, if $\kappa_s(\theta) = o(1/\varepsilon^2)$ at contact points, then $f^\varepsilon(\theta; \sigma) \to f(\theta; \sigma)$, which is defined in Eq. (2.14).

A sharp-interface model for the solid-state dewetting of a two-dimensional island/film with strongly anisotropic surface energy $\gamma(\theta)$ on a flat rigid substrate can be written as (in dimensionless form) [16, 2]:

$$(4.20) \qquad\qquad \mathbf{X}_t = \left( \mu^\varepsilon \right)_{ss} \mathbf{n}, \qquad 0 < s < L(t), \qquad t > 0.$$

The initial condition is (4.3), the first boundary condition is (i) the contact point condition (4.4), and the other boundary conditions can be written as

(ii') Relaxed contact angle condition

$$(4.21) \qquad \frac{dx_c^l}{dt} = \eta \, f^\varepsilon(\theta_d^l; \sigma), \qquad \frac{dx_c^r}{dt} = -\eta \, f^\varepsilon(\theta_d^r; \sigma), \qquad t \geq 0,$$

(iii') Zero-mass flux condition

$$(4.22) \qquad\qquad \left( \mu^\varepsilon \right)_s (0, t) = 0, \qquad \left( \mu^\varepsilon \right)_s (L, t) = 0, \qquad t \geq 0,$$

(iv) Zero-curvature condition

$$(4.23) \qquad\qquad \kappa(0, t) = 0, \qquad \kappa(L, t) = 0, \qquad t \geq 0.$$

Compared to the isotropic/weakly anisotropic case, the contact point condition (4.4) is the same and the boundary conditions (4.21)-(4.22) are similar. However, the additional boundary condition (4.23) is obtained from the first variation of the free energy (4.17) such that the system is self-closed (the dynamical evolution PDEs become sixth-order while they are fourth-order for the isotropic/weakly anisotropic case) and the total (regularized) free energy is dissipative during the evolution. In fact, this zero-curvature boundary condition (4.23) may be interpreted as the shape tends to be faceted near the two contact points when the surface energy anisotropy is strong.

As a function of time, the total (regularized) interfacial energy is defined as

$$(4.24) \qquad W_{\text{reg}}^\varepsilon(t) := \int_0^{L(t)} \left[\gamma(\theta) + \frac{\varepsilon^2}{2}\kappa^2\right] ds - \sigma[x_c^r(t) - x_c^l(t)].$$

PROPOSITION 4.2. *For the sharp-interface model governed by Eqs. (4.20) and (4.19), together with boundary conditions (4.4) and (4.21)-(4.23) and initial condition (4.3), the total area/mass is conserved (i.e. (4.8) is valid), and the total (regularized) interfacial energy decreases during the evolution, i.e.,*

$$(4.25) \qquad W_{\text{reg}}^\varepsilon(t) \leq W_{\text{reg}}^\varepsilon(t_1) \leq W_{\text{reg}}^\varepsilon(0), \quad t \geq t_1 \geq 0.$$

*Proof.* The proof that the total area/mass is conserved is similar to the proof of Proposition 4.1 and is omitted here for brevity. In order to prove that the energy is always decreasing during the evolution, the Willmore regularization energy can be rewritten as

$$(4.26) \qquad W_{\text{wm}}(t) = \frac{1}{2}\int_0^1 \kappa^2 s_p \, dp.$$

Differentiating (4.26) with respect to $t$, making use of Eq. (4.13) and the identity $\kappa_t s_p = -\theta_{pt} - \kappa s_{pt}$, applying (4.23) and integrating by parts, we obtain

$$\dot{W}_{\text{wm}}(t) = \frac{1}{2}\int_0^1 \left(2\kappa\kappa_t s_p + \kappa^2 s_{pt}\right) dp = \int_0^1 \left(-\kappa\theta_{pt} - \kappa^2 s_{pt} + \frac{1}{2}\kappa^2 s_{pt}\right) dp$$

$$= \int_0^1 \left(-\kappa\theta_{pt} - \frac{1}{2}\kappa^2 s_{pt}\right) dp = \int_0^1 \left(\kappa_s s_p \theta_t - \frac{1}{2}\kappa^2 s_{pt}\right) dp$$

$$= \int_0^1 \mathbf{X}_{pt} \cdot \left(\kappa_s \mathbf{n} - \frac{1}{2}\kappa^2 \boldsymbol{\tau}\right) dp$$

$$= -\int_0^1 \mathbf{X}_t \cdot \left(\kappa_{ss} s_p \mathbf{n} + \kappa_s \mathbf{n}_p - \kappa\kappa_s s_p \boldsymbol{\tau} - \frac{1}{2}\kappa^2 \boldsymbol{\tau}_p\right) dp + \left(\kappa_s \mathbf{X}_t \cdot \mathbf{n}\right)\Big|_{p=0}^{p=1}$$

$$(4.27) \qquad = -\int_0^{L(t)} \left(\kappa_{ss} + \frac{1}{2}\kappa^3\right)\mathbf{X}_t \cdot \mathbf{n}\, ds - \left(\kappa_s \sin\theta\right)_{\theta=\theta_d^r}\frac{dx_c^r}{dt} + \left(\kappa_s \sin\theta\right)_{\theta=\theta_d^l}\frac{dx_c^l}{dt}.$$

Combining (4.27) and (4.14), applying (4.21)-(4.22) and integrating by parts, we have

$$\dot{W}_{\text{reg}}^\varepsilon(t) = \int_0^{L(t)} \left(\widetilde{\gamma}(\theta)\kappa - \varepsilon^2\left(\kappa_{ss} + \frac{1}{2}\kappa^3\right)\right)\mathbf{X}_t \cdot \mathbf{n}\, ds + f^\varepsilon(\theta_d^r;\sigma)\frac{dx_c^r}{dt} - f^\varepsilon(\theta_d^l;\sigma)\frac{dx_c^l}{dt}$$

$$= \int_0^{L(t)} \mu^\varepsilon \left(\mu^\varepsilon\right)_{ss} ds - \frac{1}{\eta}\left[\left(\frac{dx_c^r}{dt}\right)^2 + \left(\frac{dx_c^l}{dt}\right)^2\right]$$

$$(4.28) \qquad = -\int_0^{L(t)} (\mu^\varepsilon)_s^2 \, ds - \frac{1}{\eta}\left[\left(\frac{dx_c^r}{dt}\right)^2 + \left(\frac{dx_c^l}{dt}\right)^2\right] \leq 0, \quad t \geq 0,$$

which immediately implies (4.25). □

**4.3. Non-smooth and/or 'cusped' cases.** In this case, $\gamma(\theta) \in C^0[-\pi, \pi]$ but $\gamma(\theta) \notin C^2[-\pi, \pi]$. In most anisotropic solid-state dewetting analyses, it is assumed that the surface energy $\gamma(\theta)$ is piecewise smooth and is only non-smooth and/or "cusped" at a finite set of points. A typical example can be given as [2, 32]

$$(4.29) \quad \gamma(\theta) = 1 + \beta \cos[m(\theta + \phi)] + \sum_{i=1}^{n} |\sin(\theta - \alpha_i)|, \quad \theta \in [-\pi, \pi],$$

where $n$ is a positive integer, $\beta \geq 0$ is a constant, $m$ is a positive integer, $\phi \in [0, \pi]$ and $\alpha_i \in [0, \pi]$ for $i = 1, 2, \ldots, n$ are constants.

In this situation, we can first smooth the surface energy $\gamma(\theta)$ by a $C^2$-smooth function $\gamma_\delta(\theta)$ with a smoothing parameter $0 < \delta \ll 1$ such that $\gamma_\delta(\theta)$ converges uniformly to $\gamma(\theta)$ for $\theta \in [-\pi, \pi]$ when $\delta \to 0^+$. For the above example, we use

$$(4.30) \quad \gamma_\delta(\theta) = 1 + \beta \cos[m(\theta + \phi)] + \sum_{i=1}^{n} \sqrt{\delta^2 + (1-\delta^2)\sin^2(\theta - \alpha_i)}, \quad \theta \in [-\pi, \pi].$$

If the smoothed surface energy $\gamma_\delta(\theta)$ is weakly anisotropic, $\gamma_\delta(\theta) + \gamma_\delta''(\theta) > 0$ for all $\theta \in [-\pi, \pi]$), we can apply the sharp-interface model presented in section 4.1 by replacing $\gamma(\theta)$ with $\gamma_\delta(\theta)$. On the other hand, if the smoothed surface energy is strongly anisotropic, $\gamma_\delta(\theta) + \gamma_\delta''(\theta) < 0$ for some $\theta \in [-\pi, \pi]$, we must use the (regularized) sharp-interface model with a regularization parameter $0 < \varepsilon \ll 1$ (see section 4.2) by replacing $\gamma(\theta)$ with $\gamma_\delta(\theta)$.

**5. Numerical validation.** In this section, we present six examples to validate the proposed generalized Winterbottom construction by numerically solving the proposed dynamical evolution equations for the mass transport via surface diffusion coupled with contact line migration. These examples will also show that metastable island shapes are dynamically accessible. The first four examples correspond to the four cases discussed in section 3 and the other two are for an $m$-fold smooth surface energy $\gamma(\theta)$ with a non-zero phase angle $\phi$ and non-smooth surface energy $\gamma(\theta)$ with "cusps". For the weakly anisotropic case, Eqs. (4.1)-(4.2) coupled with boundary conditions (4.4)-(4.6) are numerically solved; for the strongly anisotropic case, Eqs. (4.19)-(4.20) coupled with boundary conditions (4.4) and (4.21)-(4.23) are numerically solved. We employ a semi-implicit parametric finite element method (PFEM) recently proposed by the authors [2] for solving the above two cases. We note that for these two cases, we have rigorously proved that under the proposed models, the total area/mass of the island is conserved and the total energy decreases during the island morphology evolution. These propositions guarantee that the dynamical evolution process converges to one of stable equilibria of the total free energy functional.

In the following numerical simulations, we choose a large contact point mobility $\eta = 100$ and a small regularization parameter $\varepsilon = 0.1$ for the strongly anisotropic cases, except where noted. The choice of a large contact point mobility (e.g., $\eta = 100$) tends to drive the contact angle very quickly to the "nearest" locally-stable anisotropic Young angle determined by Eq. (2.21), resulting in an equilibrium shape that effectively minimizes the total surface energy with the fixed locally-stable contact angles. The influence of these parameter choices on the solid-state dewetting evolution process and equilibrium shapes has been discussed in [2, 16, 41].

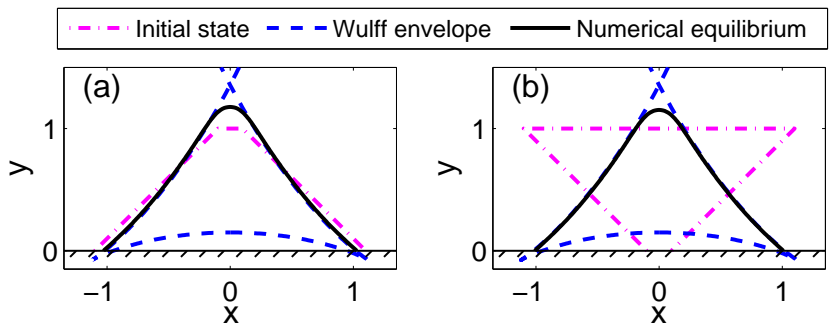

FIG. 5.1. *The numerical equilibrium shapes (solid black curves) for Case 1 in Section 3.2 with the parameters: $m = 2$, $\beta = 0.7$, $\phi = 0$, $\sigma = 1.75$ for two different initial island shapes (shown as magenta dash-dot curves). The blue dash curves represent the theoretical inverted truncated Wulff envelope obtained by the generalized Winterbottom construction.*

**Example 5.1** (*Case 1 in Section 3.2*). We set the parameters describing the anisotropic surface energy in (3.1) to $m = 2$, $\beta = 0.7$, $\phi = 0$ and $\sigma = 1.75$, as in Case 1 in Section 3.2. We choose two different initial island shapes of the same area (see the magenta dash-dot curves in Fig. 5.1(a)-(b)), and let these shapes evolve until they are stationary (*i.e.*, numerical equilibrium shapes are achieved). As shown in Fig. 5.1, for these two very different initial conditions, the numerical results converge to the same shape and this shape is that predicted from our generalized Winterbottom construction (see Fig. 3.2(d)). We note the slight differences between the theoretical and numerical result at the top corner is the result of the regularization of the surface energy. As predicted, the unstable equilibrium shape, represented by the lower blue dashed curve in Fig. 5.1 is not observed for either initial island shape.

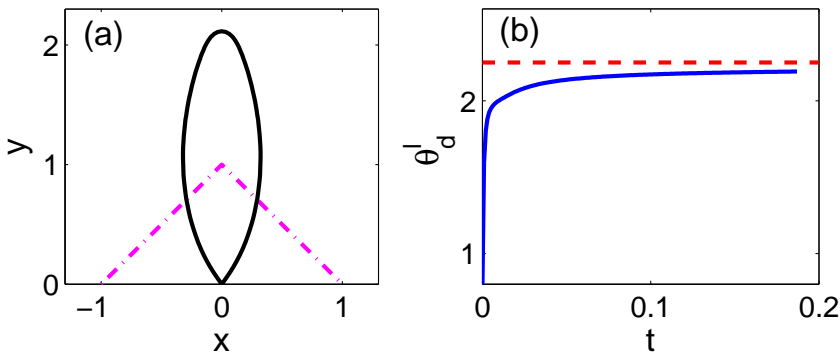

FIG. 5.2. *Simulation results for Case 2 in Section 3.2 with parameters $m = 2$, $\beta = 0.7$, $\phi = 0$, $\sigma = -1.6$: (a) shows the final state of the numerical evolution which was stopped when the two contact points meet; (b) shows the temporal evolution of the left dynamical contact angle (blue curve) as it converges to the theoretical anisotropic Young angle $\theta_a = 2.2497$ (red dash line).*

**Example 5.2** (*Case 2 in Section 3.2*). We set the parameters describing the anisotropic surface energy in (3.1) to $m = 2$, $\beta = 0.7$, $\phi = 0$, $\sigma = -1.6$. As shown in Fig. 5.2(a), we start the numerical simulation with a triangular island shape (magenta dash-dot curve). After a while, the two contact points meet - effectively decoupling the island from the substrate. This means that complete dewetting has occurred. We stop the numerical calculation when the two contact points meet, and the terminating shape of the island is shown by the solid back curve in Figure 5.2(a). This simulation result is consistent with the prediction by the Generalized Winterbottom construction shown in Fig. 3.2(e). Fig. 5.2(b) shows the corresponding temporal evolution of the left dynamic contact angle. As is clearly seen in this figure, the numerical left dynamic contact angle converges to the theoretical anisotropic Young angle $\theta_a = 2.2497$.

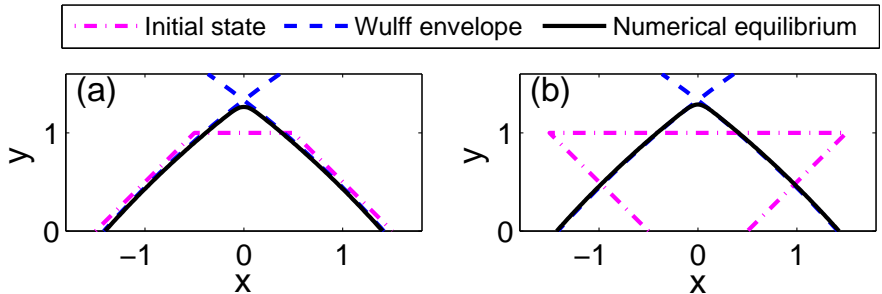

FIG. 5.3. *The numerical equilibrium shapes (solid black curves) for Case 3 in Section 3.2 with the parameters: $m = 4$, $\beta = 0.3$, $\phi = 0$, $\sigma = 0.25$ for two different initial island shapes (magenta dash-dot curves). The blue dash curve represents the theoretical truncated Wulff envelope obtained from the generalized Winterbottom construction, Fig. 3.1(c).*

**Example 5.3** (*Case 3 in Section 3.2*). We choose the same set of parameters $m = 4$, $\beta = 0.3$, $\phi = 0$, $\sigma = 0.25$ as the Case 3 in Section 3.2. As shown in Fig. 5.3, for both of the two very different initial shapes, the system evolves to the same final, stationary shape. This shape is in good agreement with the stable equilibrium shape Fig. 3.1(c) predicted by the generalized Winterbottom construction.

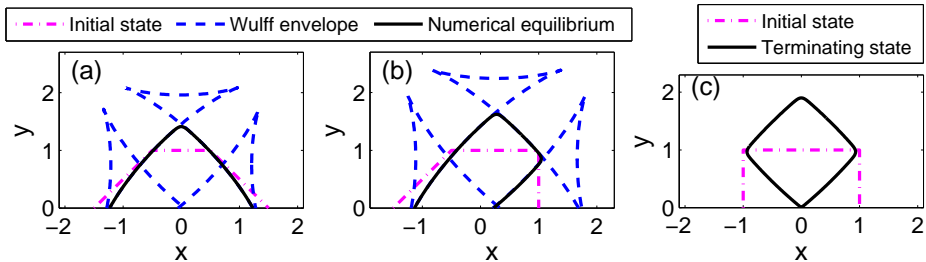

FIG. 5.4. *Multiple equilibria (black solid curves) are found for Case 4 with the parameters: $m = 4$, $\beta = 0.4$, $\phi = 0$, $\sigma = -\sqrt{3}/2$ for three different initial island shapes (magenta dash-dot curves). The energies of these three states decrease from (a) to (c), indicating that the classical Winterbottom construction prediction yields the globally minimum energy island shape (Note that the shape in (b) has a mirror symmetric counterpart, not shown here).*

**Example 5.4** (*Case 4 in Section 3.2*). For Case 4, we choose the same set of parameters $m = 4$, $\beta = 0.4$, $\phi = 0$, $\sigma = -\sqrt{3}/2$ as Fig. 3.3(b) in Section 3.2. As predicted by the generalized Winterbottom construction, three different stable equilibrium shapes and complete dewetting are the solutions. More precisely, these stable equilibria include a symmetric stable equilibrium island shape shown in Fig. 5.4(a), two asymmetric stable equilibrium shapes shown in Fig. 5.4(b) (its mirror is also a solution, not shown here); and complete dewetting shown in Fig. 5.4(c) (its numerical simulation was stopped when the two contact points meet). All of these shapes correspond to the theoretical predictions obtained from the generalized Winterbottom construction. By comparing the numerical equilibrium shapes with the generalized Winterbottom predictions, we find that they are in good accordance with each other. These results demonstrate that not only the global equilibrium island shape, but also all of the metastable island shapes are dynamically attainable. This shows that it is necessary to consider the generalization of the Winterbottom construction and not just the classical Winterbottom construction in understanding experimentally observable island morphologies.

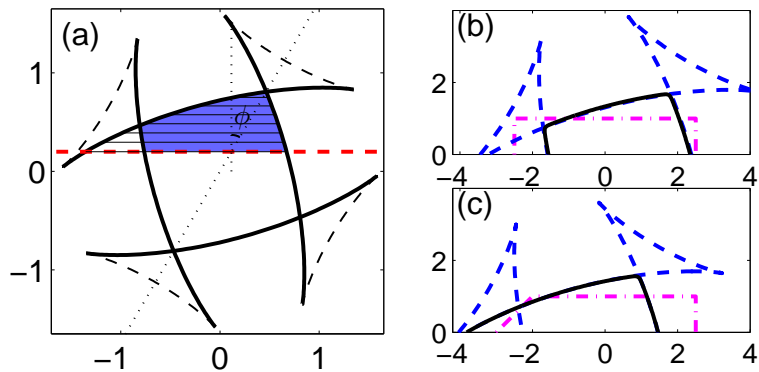

Fig. 5.5. *(a): Illustration of the generalized Winterbottom construction for an $m$-fold smooth surface energy described by the parameters $m = 4$, $\beta = 0.3$, and $\sigma = 0.2$ for the case in which the crystal structure of the island is rotated with respect to the substrate by $\phi = \pi/6$. The stable equilibrium shapes for these parameters are shown by the horizontal stripes and blue-shaded regions. (b) and (c) show the numerical equilibrium states (black solid curves) for solid-state dewetting starting with different initial shapes (magenta dash-dot curves) for the same set of parameters.*

**Example 5.5** (*$m$-fold smooth surface energy with a non-zero phase angle $\phi$*). In order to demonstrate the generality of our proposed generalized Winterbottom construction, we also present an example of an $m$-fold symmetry crystalline island with $m = 4$, $\beta = 0.3$ and $\sigma = 0.2$, where the crystal lattice of the island is rotated by $\phi = \pi/6$ (relative to the substrate surface normal). As seen in Fig. 5.5(a), two different stable equilibria are found from the generalized Winterbottom construction (blue shaded and horizontally striped regions). We show the numerically stable solutions in Fig. 5.5(b)-(c) (solid black curves) obtained for two different initial island shapes (magenta dash-dot curves). We observe that these two different initial conditions converge to two different stationary states - both of the shapes (one corresponding to the global equilibrium and one metastable) are predicted from the generalized Winterbottom construction.

**Example 5.6** (*Non-smooth surface energy with "cusps"*). For this example, we perform numerical calculations for the non-smooth (cusped) surface energy density $\gamma(\theta)$ in Eq. (4.29) for the parameters $m = 4$, $\beta = 0.25$, $\phi = 0$, $n = 1$ and $\alpha_1 = 0$. Figure 5.6(a) depicts the generalized Winterbottom construction for this non-smooth surface energy density and $\sigma = -0.3$. Applying the generalized Winterbottom construction to this case, we find four distinct stable equilibria (the asymmetric one has a mirror symmetry, not shown here). Figure 5.6(b)-(d) show the numerical equilibrium states obtained (black solid curves) starting from three different initial shapes

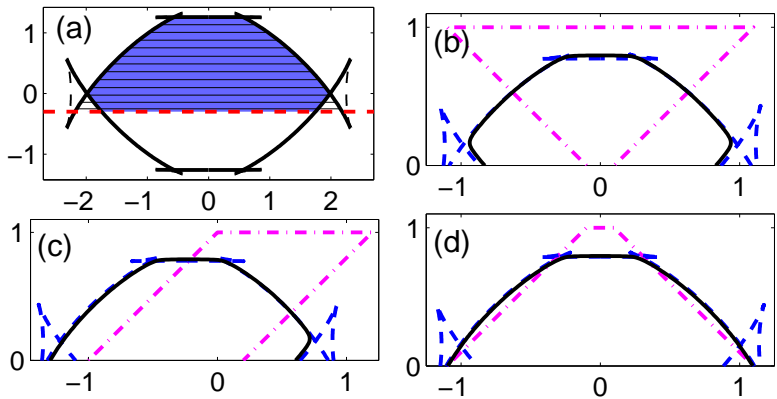

Fig. 5.6. *(a) Illustration of the generalized Winterbottom construction for a non-smooth surface energy with "cusps" as per Eq. (4.29), where the parameters were set to: $m = 4, \beta = 0.25, \phi = 0, n = 1, \alpha_1 = 0, \sigma = -0.3$. The blue shaded and horizontal striped regions indicate different stable equilibrium shapes. (b)-(d) show the numerical stable solutions (black solid curves) found from evolving the island shapes from different initial shapes (magenta dash-dot curves) using the smoothed surface energy, defined in Eq. (4.30) with $\varepsilon = \delta = 0.05$ (note that the shape in (c) has a mirror symmetric counterpart, not shown here).*

(magenta dash-dot curves) along with the Winterbottom predictions for the same set of parameters. The numerical calculations were performed with $\delta = 0.05$ in (4.30) and $\varepsilon = 0.05$ in (4.20). Again, the agreement between the numerical results and our theoretical predictions is excellent. The energies for these equilibrium shapes increase from (b) to (d) - the symmetric island shape in (b) is the classical Winterbottom prediction corresponding to the global minimum energy shape.

**6. Conclusions.** We have presented a new approach to predicting stable equilibrium shapes of two-dimensional crystalline islands on flat substrates. Our theory is a generalization of the widely used, classical Winterbottom construction [43]. This approach is equally applicable to isotropic, weakly anisotropic, strongly anisotropic and "cusped" crystal surface energy functions. We predict that, unlike in the classical Winterbottom approach, multiple equilibrium island shapes may be possible. We analyzed these shapes through a perturbation analysis, by calculating the first and second variations of the total free energy functional with respect to contact positions and island shape. Based on this analysis, we obtained the necessary conditions for the equilibria to be stable with respect to two-dimensional perturbation and exploit this through a generalization of the Winterbottom construction to identify all possible stable equilibrium shapes. Finally, we developed a dynamical evolution method based upon surface diffusion mass transport to determine if all of the stable equilibrium shapes are dynamically accessible. Applying this approach, we demonstrated that islands with different initial shapes may evolve into different stationary shapes and show that these dynamically-determined stationary states correspond to those stable, equilibrium shapes predicted by our new, generalized Winterbottom construction.